\documentclass[twocolumn,showpacs,preprintnumbers,amsmath,amssymb,prl]{revtex4}

\usepackage{graphicx}
\usepackage{dcolumn}
\usepackage{bm}

\begin{document}

\preprint{APS/123-QED}

\title{Scaling of earthquake waiting-times and the Olami-Feder-Christensen model}

\author{Morgan Hedges}
 \email{mph42@uow.edu.au}
\author{George Takacs}%
 
\affiliation{%
School of Physics, \\University of Wollongong
}%


\date{\today}

\begin{abstract}
Waiting-time statistics are generated from the Olami-Feder-Christensen model and shown to mimic some aspects of real seismicity. Preliminary analysis of the model data implies a recently proposed universal scaling law for the distribution in seismicity may be due to a mixing between aftershocks and uncorrelated event pairs, thus having limited application. Earthquake catalog data is also presented to support the argument.
\end{abstract}

\pacs{Valid PACS appear here}
\maketitle

Earthquake systems display complicated statistics in both time and space, the specific mechanisms for which remain unclear. This is despite some apparently simple power-laws describing them. The simplest, best known, and probably most significant of these laws are the Gutenberg-Richter (GR) \cite{GR} law for the probability of energy release, $P(E)\propto E^{-b}$ with $b\approx 1$, and Omori's Law for aftershocks \cite{OMORI} which states that the rate of seismic occurrence decreases as $R\propto(t+c)^{-p}$ after a large event, with $p\approx 1$. Many approaches have been considered to account for such statistics, including dynamical spring-block systems, cellular automata, and more generally the statistical frameworks of Self-Organised-Criticality (SOC) and the Epidemic Type Aftershock Sequence model (ETAS). In this paper we are concerned with a dynamical model that is generally accepted to be almost, if not precisely, SOC. The Olami-Feder-Christensen (OFC) model \cite{OFC92} is perhaps the most widely studied model of non-conservative self-organised criticality (if indeed it is technically SOC) (eg. \cite{LP01, SOCGRIN93}), and has long been known to reproduce the GR law. More recently it has also been shown to qualitatively exhibit Omori's law as well as other features of seismic aftershocks \cite{HN02,HN04}.


Correlations in time between earthquakes have long been recognised in the form of Omori clusters, but it is only recently that significant progress concerning correlation between the clusters has been made. The distribution of waiting times for a set of events, which we'll write $D(T)$, is often used to examine correlations in time. It may depend on a number of quantities, which will be discussed later. It is well known that completely uncorrelated events, describing a Poisson process, result in a pure exponential decay for $D(T)$. Variations from such indicate temporal correlations, and have long been looked for in earthquake catalogs. A short-time power law with exponent around -1 for $D(T)$ is usually taken as related to Omori's law for the rate, while there is also evidence for longer term correlations amongst large events, eg. \cite{MEGA03,KJ99}, in some cases, a faster decaying power law has been found


Following \cite{BCDS02}, Corral has proposed a ``universal'' scaling law where $D(T)$ is the same all over the world, differing only according to the seismic rate, R, in the region, so long as the rate is it is stationary. Specifically, the law states that $D(T)=R\times f(T/R)$ where $f$ is a common function holds the world over. In \cite{COR04}, Corral calculated distributions from the NEIC catalog \cite{NEIC} for many regions and periods where the rate was approximately stationary, and plotted them on a log-log graph where the axes had been rescaled by the rate. He then fit a single generalised gamma distribution to all the collapsed curves, specifically $f(\mu)=C\frac{1}{\mu^{1-\gamma}}e^{{-(\mu/\mu_0)^\delta}}$ with $\gamma=0.67\pm0.05, \delta=1.05\pm0.05, \alpha=1.64\pm0.15$. This function is essentially a decreasing power-law of exponent $1-\gamma$ giving way to an exponential decay at longer times. This fit can be seen in Fig.1 of each of \cite{COR04,COR04B}. There is a deviation from $f$ at short times, which is put down to rate fluctuations caused by aftershocks. It can be shown that a single Omori sequence should give rise to a Weibull distribution of waiting times \cite{GA04}. Although to the best of our knowledge this hasn't been confirmed in real events, we will use the result here.

In a way, all the work concerning scaling laws started from a paper by Ito, \cite{ITO95}, where the behaviour of a composite $D(T)$ constructed from dividing southern California into a grid of smaller regions was mimicked by the Bak-Sneppen model \cite{BS93}. In this way it is natural that we should look at a similar model of SOC to continue the work. We will show that the stationary rate data used by Corral may be well fit using the OFC model with certain parameters. Qualitative similarities are also found for more general behaviour. The model distributions show some complicated behaviour, but most can be described by three sections-- a short time power law of exponent approaching $-1$, a transition period, and a rapid decay, fitting qualitatively with earthquake behaviour. We show evidence that pairs of events in the distribution may be separated roughly into those within a single aftershock sequence, and those that aren't correlated. We also find long time correlations in the model in the form of a steep power law, similar to that reported in \cite{OFC92C}. We usually find and exponent of around -2.

The OFC model has been shown  \cite{OFC92} to be analagous to a 2-dimensional slider-block model (eg.\cite{CL89}) of a fault, and is described as a ``continuous time cellular-automaton''. A 2D lattice of side length $L$ is defined representing an array of sliding blocks connected to each other by springs. The blocks are imagined to be driven uniformly by an overlying plate connecting to the top of the blocks by springs. The relative stiffness of the springs between each block and the vertical springs is represented by the only other parameter of the model, $\alpha$, and governs the conservation of stress when a block slips. A value of $\alpha=0.25$ corresponds to the unphysical case of complete stress conservation, and $\alpha=0$ corresponds to complete dissipation, that is the blocks are not connected to each other at all.


The model works due to the inhomogeneity introduced by the boundaries. If not for this, as in the case of periodic boundaries, the blocks synchronise into a state where all events involve only one block. When any amount of inhomogeneity is introduced, an ordered state extrudes from there, until the whole lattice is covered in larger patches, where each block in a patch has a similar stress to it's neighbours. The patches are essential in understanding the model. They grow with distance from the boundaries, but do so much quicker with larger $\alpha$, so that their average size is strongly dependent on both $L$ and $\alpha$. The state where the whole lattice is made up of such patches we'll call the critical state, although the technical correctness of the term is debated. It is in this state that the GR and Omori's laws are obeyed. Each patch relaxes in a succession of events, and the relaxation of a single large patch appears to be the source of a foreshock/mainshock/aftershock sequence \cite{HN04}.

For our work, data was only taken after each lattice had reached the critical state, as easily checked by displaying the lattice using colour to denote the stress on each block (as in the pictures in \cite{DROS02}). We have used double precision floating point values in all simulations for which data is displayed, although we have also considered single precision in light of recent studies \cite{DROS02}, and our results are largely unmodified. We generated return-time data for lattices using open boundaries up to $L$=1024 with $\alpha=0.2$, and smaller lattices using $\alpha$ as low as $0.05$. We only present results for open boundaries here, as we found that free boundary conditions took much longer to simulate. This was due to the average event size being larger in the free case, which we put down to a longer range influence of the boundaries. A detailed analysis of mechanisms will be given in a another paper \cite{2PUB}, but the behaviour of $D(T)$ is qualitatively the same. Changing to free BCs and increasing $L$ and $M_c$ will give a similar shape to using open boundaries.
 
When taking data from the model we use the same scheme as Corral. We define a number of bins whose size increases exponentially toward longer times. When running a simulation, waiting times are placed in the appropriate bins, and the final count in each bin is divided by the bin width and the total number of events to get the average probability density for that bin. To describe magnitude in the model, we use the logarithm of the number of blocks that slipped to the base 2. This is appropriate because limitations of size in the model result in a smaller magnitude range than seen for earthquakes.

We can first make some general comments on the results as plotted with logarithmic axes. The rate of occurrence for any curve has the effect of moving a curve left or right. If events always occur in the same relative pattern, changing the unit of time should be equivalent to increasing the rate. Increasing $L$, $\alpha$ and decreasing $M_c$ all have the effect of increasing the rate. The precise value of $M_c$ is often unimportant because of the scale invariance implied by the G-R law. The effect on the rate is clear: increasing $L$ means a bigger area, and hence a higher rate. High $\alpha$ means less dissipation, so more stress remains in the lattice and more toppling occurs. Also, the general effects of $L$ and $\alpha$ may be explained in a simple manner. The size and number of the patches is the important point here-- increasing $\alpha$ increases the average patch size and hence the number of patches decreases, as indicated by a smaller $b$ value for the GR law. Decreasing $L$ has a related effect, although the effect on the $b$ value isn't as strong \cite{GRAS94, HN04}.

\begin{figure}
    \includegraphics[width=3.4in]{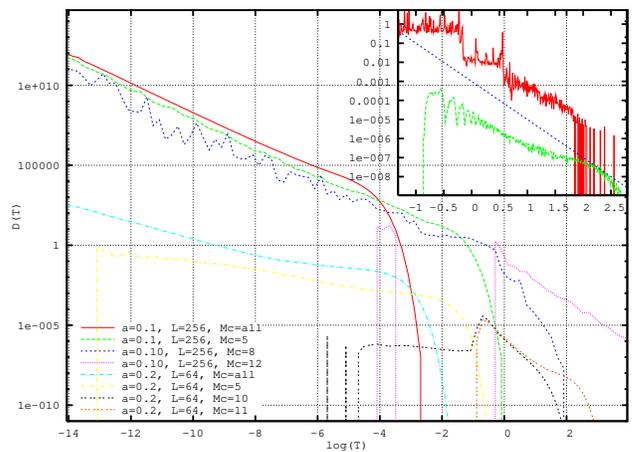}
    \caption{Results for $\alpha=0.1$, $L=256$ and $\alpha=0.2$, $L=64$ lattices with various magnitude cutoffs. The $\alpha=0.2$ curves are divided by $10^5$ for clarity. Mc for these curves refers to the cutoff magnitude defined as the logarithm with base 2 of the number of slipped blocks. The inset data was calculated using Mc=8 and 11 for each lattice respectively. The line included is a power law with exponent -2}
    \label{fig:generalOFC}
\end{figure}

Fig. \ref{fig:generalOFC} shows various $M_c$ for two contrasting lattices. The $L$=64, $\alpha=0.2$ lattice consists mainly of a single large patch, whereas the $L$=256, $\alpha=0.1$ lattice has many. There are a few features to note in this graph. First, we look at the short time, hence low $M_c$ behaviour. There is a clear short-time power law in all three $\alpha=0.1$ curves, with exponents of $\approx-0.9$.  The $\alpha=0.2$ lattice displays an exponent $\approx-0.5$ when all events are considered. When larger events are looked at only, this part of the curve begins to flatten out. We find the exponents to be consistent with the exponent for Omori's law in these lattices, using the definition d=0 as described in \cite{HN04}. This is consistent with the expected Weibull distribution of waiting times calculated from individual Omori sequences. This distribution should have a short-time power law corresponding to the Omori exponent \cite{GA04}, followed by a stretched exponential decay. If each aftershock sequence is completely defined by a single patch, then this is definitely the case. We will look for such a decay shortly.

\begin{figure}
    \includegraphics[width=3.4in]{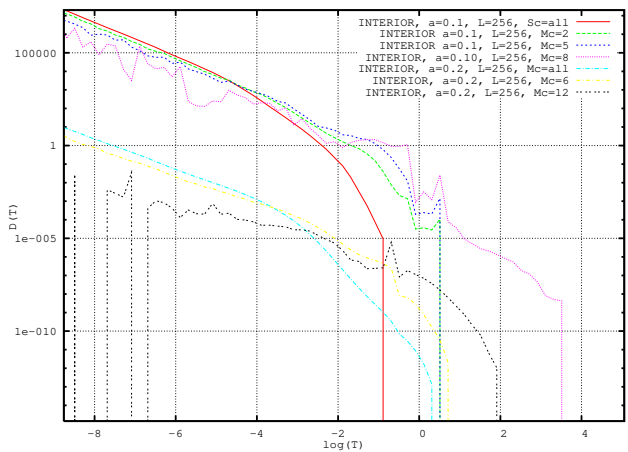}
    \caption{Results for the interior $L/2\times L/2$ section of lattices with $\alpha=0.1$, $L=256$, and $\alpha=0.2$, $L=256$. The $\alpha=0.2$ curves are divided by $10^6$ for clarity. Mc refers to the cutoff magnitude defined as the logarithm with base 2 of the number of slipped blocks. We are looking here for a decay in the distribution caused by aftershock sequences alone.}
    \label{fig:specOFC}
\end{figure}

We can see that it is appropriate to divide the graph region into two areas, divided at about $T\approx T_\alpha=1-4\alpha$. Curves passing this point show a dramatic change in behaviour. After this point, the curves in Fig.\ref{fig:generalOFC} become sharply oscillating decreasing functions, generally following a power law behaviour with exponent around -2. The inset shows this behaviour more clearly. The $\alpha=0.2$ curve is much clearer. There are actually two periods of oscillation here. The first is $\approx T_\alpha=0.2$ and the second is closer to $0.4$ This corresponds to the average period in the middle of the lattice and at the boundaries. As discussed in \cite{MB03}, $T_\alpha$ is the average period of each block in the interior in the limit of infinite lattice size. The blocks directly next to the boundaries have only three neighbours (for open BCs) and hence their average period should be $1-3\alpha$, in the middle of a side in the limit of a large lattice. As we do not use infinite lattices, we expect the period to be slightly less than this. While periodic behaviour is quite clear in  the $\alpha=0.2$ lattice, the $\alpha=0.1$ lattice shows dramatic drops at a long period interspersed with smaller scale oscillations. We think this may be due to a complicated beating behaviour due to many different periods. The oscillation occurs for large events because the patches change on a slow scale, and will relax in a similar manner on subsequent cycles. General model behaviour such as this will be discussed in a future work.

In Fig. \ref{fig:generalOFC} the $\alpha=0.1$ $L$=256, and Mc=5 curve shows a very slight indent at $T\approx10^{-3}$, although it is hard to see in that figure. The curves in Fig. \ref{fig:specOFC} where calculated only considering events initiated in the interior $L/2 \times L/2$ part of the lattice. The idea is to capture the behaviour of the larger patches far from the boundary. In these we can see that the short time power law leads into a faster, non-exponential decay without interference from the periodic behavior at $T_\alpha$. Although it is reminiscent of a power law, we consider it more likely to be related to the sum and interference of separate Weibull distributions.


Next we treat the model data in the same way as Corral treated seismic data \cite{COR04}. We include data from 5 different lattices. The first three are chosen to scale similarly to Corral's data. These lattices consist of only a single patch or so, while not being so small that they are dominated by boundary events. They show a low proportion of their events in the short-time power law. The other two are more typical, and consist of many patches. These show a large part power law behaviour. We haven't included curves which continue significantly past $T_\alpha$, as the behaviour produced prevents them from collapsing onto a single curve. Finally, we also include some curves from a lattice dominated by the boundarys. The $\alpha=0.2$, $L=32$ lattice can't form patches properly, and correlations dissapear. The exponential decay representing Poisson behaviour provides a reasonable fit, although perfect fit is impossible as the driving speed implies an inherent time-scale.

As Corral's included data is chosen for it's stationary rate, it only offers a small insight into to thee variety of seismic behaviour. To get a fuller picture, we have calculated waiting time distributions in Fig.\ref{fig:cats} for some regions of the world chosen for the goodness of their statistics, over periods for which they are complete. For this we used the combined catalog of the Advanced National Seismic Network (ANSS), \cite{ANSS}. This catalog includes the NEIC catalog used by Corral. Duplicate events are removed according to the method described on their website. It should offer a more complete source than the NEIC catalog alone.

\begin{figure}
    \includegraphics[width=3.4in]{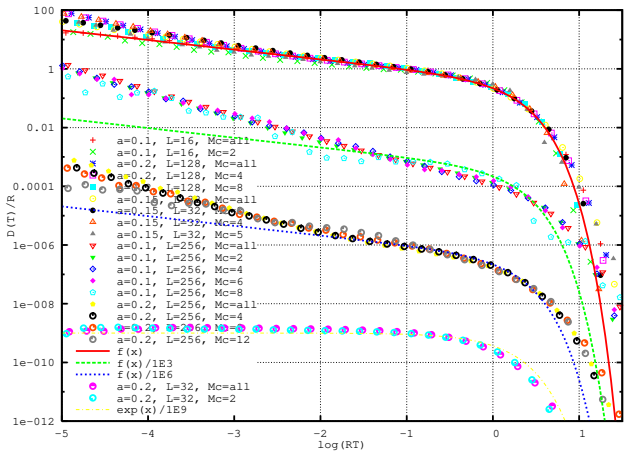}
    \caption{Various distributions plotted with axes rescaled by the rate. The top group are from lattices that give a good fit to the gamma function. These are $L=16$ $\alpha=0.1$, $L=32$ $\alpha=0.15$, and $L=128$ $\alpha=0.2$. Larger lattices or smaller $\alpha$ yields behaviour such as in the L=256 curves. Smaller $L$ or larger $\alpha$ yields random behaviour. Subsequent sets are divided by multiples of $10^3$ for clarity. Also shown for each set is $f(\mu)$, or an exponential decay for the final set representing a Poissonian result. }
    \label{fig:scalOFC}
\end{figure}

\begin{figure}
    \includegraphics[width=3.4in]{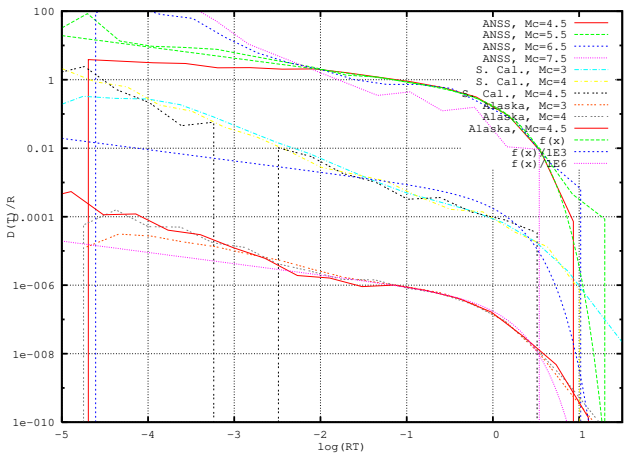}
    \caption{Various distributions calculated from the ANSS worldwide catalog. The regions and periods shown are the world 1973-2004, Southern California 1984-2004, and Alaska 1971-2004. The periods are chosen based on the catalog submission details page of the ANSS website, \cite{ANSS}. The regions are also defined there. The Southern Californian and Alaskan curves are divided by $10^3$ and $10^6$ respectively.}
    \label{fig:cats}
\end{figure}

The first curves are for events anywhere in the world. We can see that for lower cutoffs the scaling law is very well satisfied, but for larger cutoffs the short-time power law behaviour increases thus deviating at short times, while there is also a deviation from the exponential decay at longer times. The curves calculated from the Southern California and Alaskan regions of the ANSS catalog look quite similar to some of the curves in Fig. \ref{fig:scalOFC}. The behaviour of these catalogs we've found to be representative of many highly active regions of the world.

To understand the scaling behaviour, we first note that the Poisson distribution is described solely by the rate, and gives pure exponential waiting-time distribution. Because it depends only on the rate, scaling the axes it's plotted against in this way results simply in $e^{-x}$ every time. In the model curves that scale well, the majority of events are in the exponential tail, so it is no surprise that they move together. As the number of aftershocks becomes significant, the power-law behaviour becomes significant resulting in poorer scaling. The model curves that scale well have the common property of consisting of only a few patches, and thus they have similar relative aftershock activity. The right amount of deviation from uncorrelated behaviour seems to be required to fit the gamma function, and it is not a general rule.

As for earthquakes, we must work with a smaller number of events, and definite conclusions are more difficult. We have shown however that when general behaviour is looked at, without isolating specific periods of stationarity, $f$ is the exception rather than the rule as in the model. Further, when we look at the ANSS catalog data we see that the $f$ is only well fit when a cutoff magnitude of about 5.5 is used. Lower gives a more random appearing curve, while higher gives more aftershock behaviour. This implies that the goodness of fit may be due to either incompleteness in the catalog, or simply a lack of correlations resulting in more randomness, when smaller events are considered.

We have presented waiting-time statistics generated from the OFC model with open boundary conditions, showing that the general shape and scaling behaviour of real earthquake distributions is reproduced. It appears that the generated distributions are formed from two regimes and so universal behaviour is not found. The reason for the scaling and the gamma function fit are then a majority of Poisson-like events, and a transition region between the regimes which approximates a slow decay power-law. Given the available evidence, it seems likely that real events scale for the same reason-- when aftershock activity is low, that is the rate is approximately constant, most events follow a Poisson distribution. Also of note, we find that the waiting times of many large events describe an $\approx-2$ exponent power law decay as has sometimes been found earthquakes. It is tempting to hypothesise that earthquakes may show this behaviour for a similar reason: that large areas of the earth remain in an organised state that changes only slowly with time. In this case, a characteristic time governed by the slow driving would be apparent. In the model, it is the time taken for a block to go from zero stress to the threshold influenced only by the slow driving. For earthquakes, it could be the time taken for the friction in a single fault to build to it's slipping value, without being influenced by others.

We should note that for our comparisons to real events, we are interpreting the model as representing a whole earthquake system, although it was originally intended to represent a single fault. This is however in line with the treatment in \cite{LP02} on the fractal distribution of epicenters. In this case, the stress transfer may correspond roughly with the transfer of stress between sections of fault, although the analogy is somewhat lacking. The OFC model as it stands is a very unrealistic model of an entire seismic system for a number of reasons, the main one being that real systems have no well defined boundaries, and are inhomogeneous across their area. The value of our results is perhaps not in revealing the fundamental behaviour of seismicity, but a rather more humble exploration of the mechanisms that may go in to constructing the often confusing waiting-time distribution. Further work will is required to determine if a real link to seismicity exists.


\bibliography{refs}

\end{document}